\documentclass[final,5p,times]{elsarticle}
\usepackage{amssymb,bbold}
\usepackage{natbib}
\usepackage{hyperref}
\usepackage[latin1]{inputenc}
\usepackage{float}
\usepackage{graphicx}
\usepackage{graphics}
\usepackage{amsfonts}
\usepackage{color}
\usepackage[T1]{fontenc}
\usepackage[english]{babel}
\usepackage{wrapfig}
\usepackage{amsmath}
\usepackage{amssymb}
\usepackage{mathrsfs}
\usepackage{subfigure}
\hypersetup{
  colorlinks=true,linkcolor=blue,citecolor=blue,
  filecolor=blue,urlcolor=blue,breaklinks=true
}
\biboptions{sort&compress}
\journal{Physics Letters A}
\begin{document}

\begin{frontmatter}

\title{Influence of spatially varying pseudo-magnetic field on a 2D electron gas in graphene}

\author[ufcg]{L. G. da Silva Leite}
\ead{lazaroleite@hotmail.com.br}
\author[ufcg]{C. Filgueiras}
\ead{cleversonfilgueiras@yahoo.com.br}
\author[ufcg]{D. Cogollo}
\ead{diegocogollo@df.ufcg.edu.br}
\author[ufma]{Edilberto O. Silva}
\ead{edilbertoos@pq.cnpq.br}

\address[ufcg]{Unidade Acad\^{e}mica de F\'{\i}sica, Universidade Federal de Campina Grande, Caixa Postal 10071, 58109-970 Campina Grande, Para\'{\i}ba, Brazil}

\address[ufma]{Departamento de F\'{\i}sica, Universidade Federal do Maranh\~{a}o, Campus Universit\'{a}rio do Bacanga, 65085-580 S\~{a}o Lu\'{\i}s-MA, Brazil}

\begin{abstract}
The effect of a varying pseudo-magnetic field, which falls as $1/x^2$, on a two dimensional electron gas in graphene is investigated. By considering the second order Dirac equation, we show that its correct general solution is that which might present singular wavefunctions since such field induced by elastic deformations diverges as $x\rightarrow0$. We show that only this consideration yields the known relativistic Landau levels when we remove such elastic field. We have observed that the zero Landau level fails to develop for certain values of it.
We then speculate about the consequences of these facts to the quantum Hall effect on graphene. We also analyze the changes in the relativistic cyclotron frequency. We hope our work being probed in these contexts, since graphene has great potential for electronic applications.
\end{abstract}


\end{frontmatter}

\section{Introduction}

In 2004, the discovery of an one atom thick material was announced,
which rapidly caught the attention of many physicists \cite{novo}. Graphene,
a single layer of carbon atoms in a honeycomb lattice, is considered a truly
two dimensional system. The carriers within it behave as two-dimensional
massless Dirac fermions \cite{review}. Due to its peculiar physical
properties, graphene has great potential for nanoelectronic applications
\cite{rev1,nature,capac}. Graphene can be considered a zero-gap
semiconductor. This fact prevents the pinch off of charge currents
in electronic devices. Quantum confinement of electrons and holes in
nanoribbons \cite{rib} and quantum dots \cite{dot} can be realized in order
to induce a gap. However, this lattice disorder suppresses an
efficient charge transport \cite{mucc,naka}. One alternative to open a gap
is to induce a strain field in a graphene sheet onto appropriate substrates
\cite{cocco}. They play the role of an effective gauge field which yields a
pseudo-magnetic field \cite{nature1}. Unlike actual magnetic
fields, these strain induced pseudo-magnetic fields do not violate
the time reversal symmetry \cite{reversal,weh}.

Recently, some works devoted to the search for solutions of the Dirac
equation with position dependent magnetic fields were addressed \cite%
{vary,JP.CM,sec,p1}. However, they consider \textit{actual} instead of
\textit{pseudo-magnetic} fields. No experiments have been reported yet and
we believe it is because such field configurations are not easy to implement
in the laboratory. In this paper, we investigate a graphene sheet in the
presence of both a constant orthogonal magnetic and an orthogonal
pseudo-magnetic field. We consider the pseudo-magnetic field
falling as $1/x^{2}$. This configuration is not known experimentally, but in
considering it as induced by elastic deformations in graphene, we believe
someone would be able to implement it in the laboratory. Moreover, this is
the simplest case where we can get analytical solutions.
Specifically, we investigate how such non constant pseudo-magnetic
field modifies the relativistic Landau levels. We will solve the squared
Dirac equation and show that among the possible choices for the
wavefunction, the correct is the one that diverges at the origin of
the coordinate system. This is compatible with the fact that our
differential equation diverges at the origin as well. In Ref. \cite{comment}%
, it is discussed that this is the correct choice if singularity is
taken into account. Otherwise, we would get the wrong spectrum. This is
also in agreement with other quantum problems where singularities have also
appeared. This question about the correct behavior of wavefunctions
whenever we have singularities has been investigated via the \textit{self
adjoint extension approach} over the last years \cite{self}. An important
result is that the zero-energy, which exist in the known relativistic Landau levels when just the constant orthogonal magnetic field is present, does not show up for a specific range of the parameter characterizing the varying
pseudo-magnetic field. The consequence is that a Hall plateau
develops at the null \textit{filling } factor
(dimensionless ratio between the number of charge carriers and the
flux quanta). Modifications in the \textit{relativistic cyclotron frequency} are examined as well.

\section{Relativistic Landau levels}

In this section, we will investigate how a varying pseudo-magnetic
field perpendicular to a graphene sheet is going to affect the relativistic
Landau levels. First, we must remember the reader that the low-energy excitations of graphene behave as massless Dirac fermions, instead of
massive electrons. These low-energy excitations are described by
the $(2+1)$-dimensional Dirac equation
\begin{equation}
-iv_{F}\left( \mathbf{\sigma}\cdot \mathbf{\nabla }\right) \Psi (\mathbf{r}
)=E\Psi (\mathbf{r}),  \label{eq:dirac}
\end{equation}%
where $\sigma =\left( \sigma _{x},\sigma _{y}\right) $ are the Pauli
matrices, $\Psi =(\varphi _{1},\varphi _{2})^{T}$ is a two-component
spinor field, the speed of light $c$ was replaced by the Fermi
velocity ($v_{F}\approx 10^{6}$m/s) and $\hbar $ has been fixed equal to
one. The electronic states around the zero energy are states belonging to
distinct sublattices. This is the reason we have a two component
wavefunction. Two indexes to indicate these sublattices, similar to spin
indexes (up and down), must be used. The inequivalent cornes of the
Brillouin zone, which are called \textit{Dirac points}, are labeled as $K$
and $K^{^{\prime }}$ \cite{etrans,geim}.

In this work, the varying pseudo-magnetic field is supposed to
appear due to strains on a graphene sheet \cite{strain1}. The
valleys $K$ and $K^{^{\prime }}$ feel an effective field of $\tilde{\mathbf{A%
}}\pm \mathbf{A}$, where $\tilde{\mathbf{A}}$ is due to a real magnetic
field and $\mathbf{A}$ is due to a pseudo-magnetic field. Notice
that a different sign has to be used for the gauge field due to strain at
the valleys $K$ and $K^{^{\prime }}$ since such fields do not break time
reversal symmetry \cite{strain2}. Considering the Landau gauge, we have
\begin{equation}
\tilde{A}\pm A=\left[ A_{x}=0,A_{y}=\left( B_{0}x\pm \frac{\lambda }{x}%
\right) ,A_{z}=0\right] \;,
\end{equation}%
where $\lambda $ is a constant. This way, the magnetic field is $\mathbf{B}=%
\left[ B_{0}\pm \frac{\lambda }{x^{2}}\right] \mathbf{\hat{z}}$. The first
term in this field, $B_{0}$, corresponds to a constant magnetic field
along the $z$ direction which is perpendicular to the graphene
plane.

Going back to the problem, we consider the electronic states around the
valley $K$ and the minimal coupling for electrons as $-i\mathbf{\nabla }%
\longrightarrow -i\mathbf{\nabla }+e\mathbf{A}=\mathbf{\pi }$. Then,
\begin{equation}
\pi =\left[ p_{x},p_{y}+e\left( B_{0}x-\frac{\lambda }{x}\right) \right] .
\label{cou}
\end{equation}%
For the valley $K^{^{\prime }}$ we make the change $\lambda \rightarrow
-\lambda $.

The Hamiltonian is given by
\begin{equation}
H=v_{F}\left( \mathbf{\sigma }\cdot \mathbf{\pi }\right) .
\end{equation}%
Writing
\begin{equation}
\mathbf{\sigma }\cdot \mathbf{\pi }=\sigma _{x}\pi _{x}+\sigma _{y}\pi _{y},
\end{equation}%
and using Eq. (\ref{cou}), we can write Eq. (\ref{eq:dirac}) in the form
\begin{equation}
\begin{pmatrix}
\dot{\varphi _{1}} \\
-\dot{\varphi _{2}}%
\end{pmatrix}%
=%
\begin{pmatrix}
0 & W \\
Z & 0%
\end{pmatrix}%
\begin{pmatrix}
\varphi _{1} \\
\varphi _{2}%
\end{pmatrix}%
,  \label{matrix}
\end{equation}%
where
\begin{equation}
W=-v_{F}\partial _{x}+iv_{F}\left[ \partial _{y}+ie\left( B_{0}x-\frac{%
\lambda }{x}\right) \right] ,
\end{equation}%
and
\begin{equation}
Z=v_{F}\partial _{x}+iv_{F}\left[ \partial _{y}+ie\left( B_{0}x-\frac{%
\lambda }{x}\right) \right] .
\end{equation}%
Taking the time derivative of Eq. (\ref{matrix}), we are able to write the
second order Dirac equations for both $\varphi _{1}$ and $\varphi _{2}$,
that is
\begin{equation}
\ddot{\varphi _{1}}=-WZ\varphi _{1},
\end{equation}%
and
\begin{equation}
\ddot{\varphi _{2}}=-ZW\varphi _{2}.  \label{c}
\end{equation}%
Considering
\begin{equation}
\varphi _{2}(\mathbf{r},t)=e^{-iEt}\varphi (\mathbf{r}),
\end{equation}%
where $\varphi (\mathbf{r})$ is the spatial part of the spinor component $%
\varphi _{2}(\mathbf{r},t)$, we obtain
\begin{eqnarray}
&&{E^{2}\varphi (\mathbf{r})}=\left[ v_{F}\partial _{x}+iv_{F}\partial
_{y}-v_{F}eB_{0}x+v_{F}e\frac{\lambda }{x}\right]   \notag \\
&&\times \left[ -v_{F}\partial _{x}+iv_{F}\partial _{y}-v_{F}eB_{0}x+v_{F}e%
\frac{\lambda }{x}\right] \varphi (\mathbf{r}).  \label{e1}
\end{eqnarray}%
Equation (\ref{e1}) above provides
\begin{eqnarray}
&&E^{2}\varphi (\mathbf{r})=v_{F}^{2}\Big[-\nabla
^{2}-eB_{0}-2e^{2}B_{0}\lambda +\frac{e\lambda \left( e\lambda -1\right) }{%
x^{2}}  \notag \\
&&+2ie\frac{\lambda }{x}\partial _{y}-2ieB_{0}x\partial
_{y}+e^{2}B_{0}^{2}x^{2}\Big]\varphi (\mathbf{r}).  \label{d}
\end{eqnarray}%
The wavefunction can be factorized as $f(x)g(y)$. Since the vector potential
depends on the $x$ coordinate, only the fermions will
behave as plane waves in the $y$ direction. Then, we consider the \textit{%
ansatz} for (\ref{d}) as
\begin{equation}
\varphi (\mathbf{r})=f(x)e^{ik_{y}y},
\end{equation}%
which yields
\begin{eqnarray}
&&\frac{d^{2}f(x)}{dx^{2}}+\Big[\epsilon -\frac{e\lambda \left( e\lambda
-1\right) }{x^{2}}+\frac{2ek_{y}\lambda }{x}-2eK_{y}B_{0}x  \notag \\
&&-e^{2}B_{0}^{2}x^{2}\Big]f(x)=0,  \label{dif1}
\end{eqnarray}%
where
\begin{equation}
\epsilon =\frac{E^{2}}{v_{F}^{2}}-k_{y}^{2}+2e^{2}B_{0}\lambda +eB_{0}.
\end{equation}%
By defining the dimensionless variable
\begin{equation}
\chi =\sqrt{e\left\vert B_{0}\right\vert }x,
\end{equation}%
Eq. (\ref{dif1}) reads%
\begin{equation}
f^{\prime \prime }(\chi )+\left[ \mathrm{C}+\frac{\mathrm{F}}{\chi ^{2}}+%
\frac{\mathrm{D}}{\chi }+\mathrm{B}\chi -\chi ^{2}\right] f(\chi )=0,
\label{f}
\end{equation}%
where
\begin{eqnarray}
\mathrm{B} &=&-2k_{y}\sqrt{1/(e\left\vert B_{0}\right\vert )}\frac{B_{0}}{%
\left\vert B_{0}\right\vert },  \notag \\
\mathrm{C} &=&\frac{\epsilon }{e\left\vert B_{0}\right\vert },  \notag \\
\mathrm{D} &=&2ek_{y}\lambda \sqrt{1/(e\left\vert B_{0}\right\vert )},
\notag \\
\mathrm{F} &=&-e\lambda \left( e\lambda -1\right) .  \label{using}
\end{eqnarray}%
The general solution of this differential equation can be obtained by using
the Frobenius method to find series expansions. A similar differential
equation was obatined in Ref. \cite{eug} and it was found that
\begin{equation}
f(\chi )=\left\vert \chi \right\vert ^{\beta }e^{-\alpha \chi ^{2}-\gamma
\chi }\mathrm{HeunB}(\chi ),  \label{ansatz}
\end{equation}%
where $\beta $, $\alpha $, $\gamma $ are constants and $\mathrm{HeunB}$ is
the so called \textit{biconfluent Heun function }\cite{heun}. We considered
the modulus in the first piece in (\ref{ansatz}) since $\chi \in \left(
-\infty ,\infty \right) $ \cite{cone}. By substituting Eq. (\ref{ansatz})
into Eq. (\ref{f}), it results in
\begin{eqnarray}
&&f\left( \chi \right) ={C_{1}}\,{\left\vert \chi \right\vert }^{\frac{1}{2}%
\left( 1+\sqrt{1-4\mathrm{\,F}}\right) }\mathrm{e}^{-\frac{1}{2}\left( \chi
^{2}-\mathrm{B\chi }\right) }  \notag \\
&&\times \,\mathrm{HeunB}\left( \sqrt{\mathrm{1-4\,F}},\mathrm{B},\mathrm{C}+%
\frac{1}{4}\,\mathrm{B}^{2},2\,\mathrm{D},-\chi \right)   \notag \\
&&+\,{C_{2}}\,{\left\vert \chi \right\vert }^{\frac{1}{2}\left( 1-\sqrt{1-4%
\mathrm{\,F}}\right) }\mathrm{e}^{-\frac{1}{2}\left( \chi ^{2}-\mathrm{B\chi
}\right) }  \notag \\
&&\times \,\mathrm{HeunB}\left( -\sqrt{\mathrm{1-4\,F}},\mathrm{B},\mathrm{C}%
+\frac{1}{4}\mathrm{B}^{2},2\,\mathrm{D},-\chi \right) ,  \label{general}
\end{eqnarray}%
where $C_{1}$ and $C_{2}$ are normalization constants. In order to
investigate bound states, the general wavefunction must be
square-integrable:
\begin{equation}
\int_{-\infty }^{\infty }\left\vert f\left( \chi \right) \right\vert
^{2}d\chi <\infty .  \label{eq:1}
\end{equation}%
We then analyze the asymptotic behavior of solutions to the equation above,
for $\chi \rightarrow 0^{\pm }$ and $\chi \rightarrow \pm \infty $. The
exponential term guarantees that
\begin{equation}
\lim_{\chi \rightarrow \pm \infty }f\left( \chi \right) \rightarrow 0,
\end{equation}%
if the series $\mathrm{HeunB}$ reduces to a polynomial of degree $n$. No
further condition must be considered for the wavefunction. But we must be
careful in choosing the right solution since our differential equation (\ref%
{f}) has a singularity at $\chi =0$. Considering $C_{1}\equiv 0$ in Eq. (\ref%
{general}), we have
\begin{eqnarray}
&&\lim_{\chi \rightarrow 0^{\pm }}\,{\left\vert \chi \right\vert }^{\frac{1}{%
2}\left( 1+\sqrt{1-4\mathrm{\,F}}\right) }\mathrm{e}^{-\frac{1}{2}\left(
\chi ^{2}-\mathrm{B}\chi \right) }  \notag \\
&&\times \mathrm{HeunB}\left( \sqrt{1-4\,\mathrm{F}},\mathrm{B},\mathrm{C}+%
\frac{1}{4}\,\mathrm{B}^{2},2\,\mathrm{D},-\chi \right) \mathrm{\rightarrow
0,}
\end{eqnarray}%
which means that the wavefunction is regular at the origin ($\mathrm{%
HeunB\rightarrow 1}$ as $\chi \rightarrow 0$ \cite{heun}). On the other
hand, if we take $C_{2}\equiv 0$, the wavefunction (\ref{general})
might diverge at the origin because of the term ${\left\vert \chi
\right\vert }^{\left( 1-\sqrt{1-4\,\mathrm{F}}\right) /2}$. This said, we
conclude that the solution compatible with the fact that our differential
equation is singular at $\chi =0$ must be that with $C_{2}\equiv 0$ in Eq. (%
\ref{general}), i.e., we put $C_{1}\equiv 0$, and
\begin{eqnarray}
&&f(\chi )={C_{2}}\,{\left\vert \chi \right\vert }^{\frac{1}{2}\left( 1-%
\sqrt{1-4\mathrm{\,F}}\right) }\mathrm{e}^{-\frac{1}{2}\left( \chi ^{2}-%
\mathrm{B\chi }\right) }  \notag \\
&&\times \,\mathrm{HeunB}\left( -\sqrt{1-4\,\mathrm{F}},\mathrm{B},\mathrm{C}%
+\frac{1}{4}\,\mathrm{B}^{2},2\,\mathrm{D},-\chi \right) .  \label{sol}
\end{eqnarray}%
Otherwise, we can get the wrong spectrum as discussed in \cite{comment}. We
will see bellow that this is the only choice which recovers the
known spectrum for an orthogonal constant magnetic field, $B_{0}$. Notice
that a divergence in the wavefunction happens when
\begin{equation}
\frac{1}{2}\left( 1-\sqrt{1-4\mathrm{\,F}}\right) <0.  \label{E1}
\end{equation}%
From this last equation, the divergence in the wavefunction exists for $%
e\lambda <0$ and for $e\lambda >1$.

\begin{figure}[th]
\centering
\includegraphics[height=6cm]{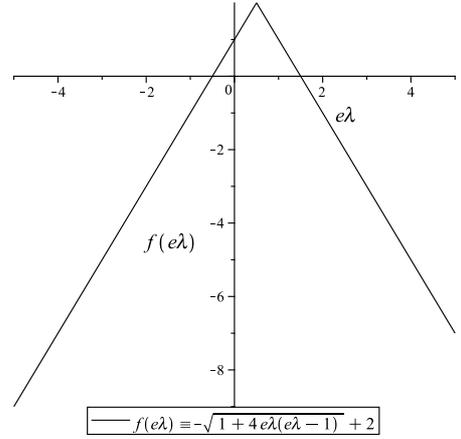}
\caption{This plot shows that bound states exist for $-1/2<e\protect\lambda %
<3/2$ since the quantum probability does not diverge in this interval.}
\label{fig1}
\end{figure}

We must investigate the behavior of quantum probability as $\chi \rightarrow
0$. When $\mathrm{HeunB}$ is a polynomial of degree $n$, we have
\begin{eqnarray}
\lim_{\chi \rightarrow 0^{\pm }}\int \left\vert f\left( \chi \right)
\right\vert ^{2}d\chi  &=&{\left\vert C_{2}\right\vert ^{2}}\lim_{\chi
\rightarrow 0^{\pm }}\int \left[ \left\vert \chi \right\vert ^{\frac{1}{2}%
\left( 1-\sqrt{1-4\mathrm{\,F}}\right) }\right] ^{2}d\chi ,  \notag \\
&=&\mathrm{constant}\times \left\vert \chi \right\vert ^{-\sqrt{1-4\mathrm{%
\,F}}+2}.  \label{eq:2}
\end{eqnarray}%
To avoid divergence in this equation, we must impose
\begin{equation}
-\sqrt{1-4\mathrm{\,F}}+2>0.  \label{E2}
\end{equation}%
In Fig. \ref{fig1}, it is depicted Eq. (\ref{E2}). It shows that the
parameter $e\lambda $ can assume any real value in the interval $(-1/2,3/2)$%
. This interval comes from finding the roots of Eq. (\ref{E2}). Before
continuing, we must mention that the $1/x^{2}$ potential can lead
to the "fall to the center" problem \cite
{1x2}. In order to prevent this phenomenon, we must have $1-4\mathrm{F}\geq 0
$. This expression can be put in two forms, $(1-2e\lambda )^{2}$ and $%
(-1+2e\lambda )^{2}$. We conclude that physical solutions appear in the
interval $(-1/2,3/2)$. For $e\lambda =1/2$, we have $1-4\mathrm{F}=0$. As we
saw above, the wavefunctions diverge for $e\lambda <0$ and $e\lambda >1$.
This means that regular solutions exist for $0\leq e\lambda \leq 1$ and
irregular solutions exist for $-1/2<e\lambda <0$ and $1<e\lambda <3/2$.
Finally, the biconfluent Heun series becomes a polynomial of degree $n$ when
\cite{heunB}
\begin{equation}
C+\frac{1}{4}\mathrm{B}^{2}=2n+2-\sqrt{1-4\mathrm{\,F}},  \label{ser}
\end{equation}%
with $n=0,1,2,3...$. Putting $E\equiv E_{n}^{\lambda }$ and using Eq. (\ref%
{using}), we arrive at
\begin{equation}
E_{n}^{\lambda }=\pm v_{F}\sqrt{2e\left\vert B_{0}\right\vert \left( n+\frac{%
1}{2}-\frac{\sqrt{1+4e\lambda (e\lambda -1)}}{2}-e\lambda \right) }.
\label{landau}
\end{equation}%
Notice that, for $\lambda =0$, we get
\begin{equation}
E_{n}^{0}=\pm v_{F}\sqrt{2e\left\vert B_{0}\right\vert n},  \label{en}
\end{equation}%
which is the known relativistic Landau levels expression for
massless fermions in the presence of a constant orthogonal magnetic field.
If we had chosen the regular wavefunction in Eq. (\ref{general}), we would
have found $E_{n}^{0}=\pm v_{F}\sqrt{2e\left\vert B_{0}\right\vert \left(
n+1\right) }$ as $\lambda \rightarrow 0$. So, this corroborates with our
statement above that the correct solution must be that which may show
divergence in the wavefunction at the origin.
\begin{figure}[th]
\begin{center}
\includegraphics[height=7cm]{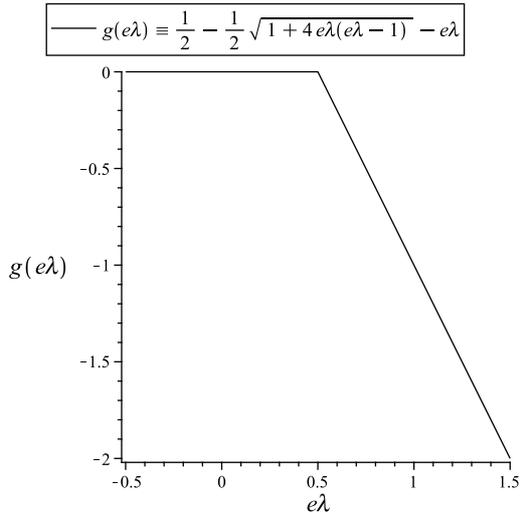}
\end{center}
\caption{{\protect\scriptsize Plot of $g(e\protect\lambda )=(E_{0}^{\protect%
\lambda })^{2}/2ev_{F}^{2}\left\vert B_{o}\right\vert $ versus $e\protect%
\lambda $. From it, we can see that this zero mode does not exist for
certain $e\protect\lambda $(the eigenvalues are imaginary in this region).}}
\label{fig2}
\end{figure}
It is useful to plot the energy versus the parameter $e\lambda $ in order to
see clearly the modifications introduced by this varying
pseudo-magnetic field. As it is known, there is a zero energy mode
for $n=0$ in Eq. (\ref{en}). In Fig. \ref{fig2}, we plot $%
(E_{0}^{\lambda })^{2}/2ev_{F}^{2}\left\vert B_{0}\right\vert $ for the $n=0$
mode. From it, we observe that the zero mode does not show up when $%
1/2<e\lambda <3/2$ since for $n=0$ the eigenvalues are imaginary.
The zero mode still exists for $-1/2<e\lambda \leq 1/2$. We now look to the
anomalous quantum Hall effect on graphene to see the consequence of this
result. The Hall conductivity is generally given by $\sigma _{xy}=\nu e^{2}/h
$, where $\nu $ is the filling factor (dimensionless ratio between
the number of charge carries and the flux quanta), $e$ is the electrical
charge and $h$ is the Planck$^{,}$s constant. At the Dirac point, both holes
and electrons coexist at the zero energy and there is a finite (and
quantized) contribution to the transverse conductivity given by $\pm 2e^{2}/h
$. In simple words, varying the concentration of charge carries the
Hall conductivity $\sigma _{xy}$ will show up as an uninterrupted ladder of
equidistant steps \cite{review,naturemat}. Ignoring the many-body
effects, the Hall conductivity on graphene is given by $\sigma _{xy}=\pm
4e^{2}/h(n+1/2)$, where $n$ is the Landau level index and the factor
4 appears due to double valley and double spin degeneracy. This expression
shows that plateaus of conductivity are formed when $\nu =\pm 4(n+1/2)=\pm
2,\pm 6,\pm 10...$. The filling factors $\nu =\pm 2$ correspond to the $n=0$ mode (zero energy). When we turn on the varying magnetic field, the
zero Landau level fails to develop and the plateaus at $\nu =\pm 2$ collapse
into one single plateau at $\nu =0$ \cite{split}. Then, we have the
subsequent plateaus formerly at $\nu =\pm 6$ appearing at $\nu =\pm 4$, and
so on. Further analysis about the quantum Hall effect, taking into account
the electron-electron interactions \cite{p2,p3,indian}, should be carried
out in a future work.
\begin{figure}[th]
\begin{center}
\includegraphics[height=7cm]{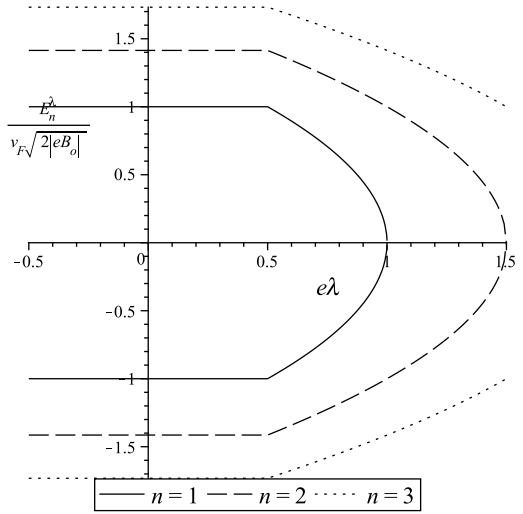}
\end{center}
\caption{{\protect\scriptsize Plot of energy versus the parameter $e\protect%
\lambda $. As we can see, the positive energies (holes) diminish and the
negative energies (electrons) in the interval $1/2<e\protect\lambda <3/2$.
They are unchanged when $-1/2<e\protect\lambda <1/2$.}}
\label{landaulp}
\end{figure}

\begin{figure}[th]
\begin{center}
\includegraphics[height=7cm]{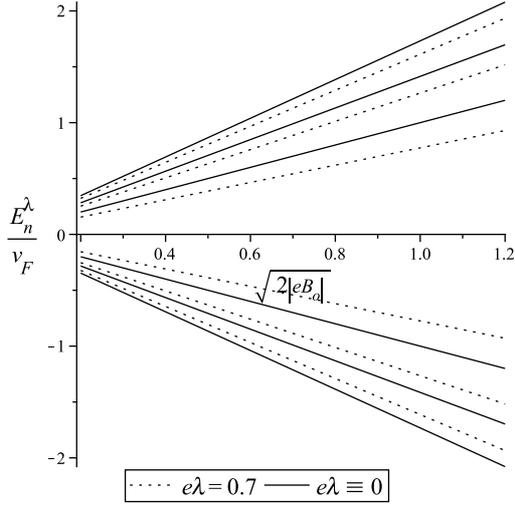}
\end{center}
\caption{{\protect\scriptsize Plot of energy versus $\protect\sqrt{%
2\left\vert eB{_{o}}\right\vert }$ for $n=1,2,3$. The energies shift to
lower values for holes (upper curves) and to higher values for electrons
(lower curves). The $e\protect\lambda =0$ case corresponds to the
relativistic Landau levels for a constant orthogonal magnetic field alone.}}
\label{fig5}
\end{figure}

In Fig. \ref{landaulp}, we plot the Landau levels (\ref{landau}) versus the
parameter $e\lambda $ for $n=1,2,3$. The energies shift to lower values for
positive energies (holes) and to higher values for negative energies
(electrons) when $1/2<e\lambda <3/2$ (see Fig. \ref{fig5}).
\begin{figure}[!t]
\centering
a)\includegraphics[height=8cm]{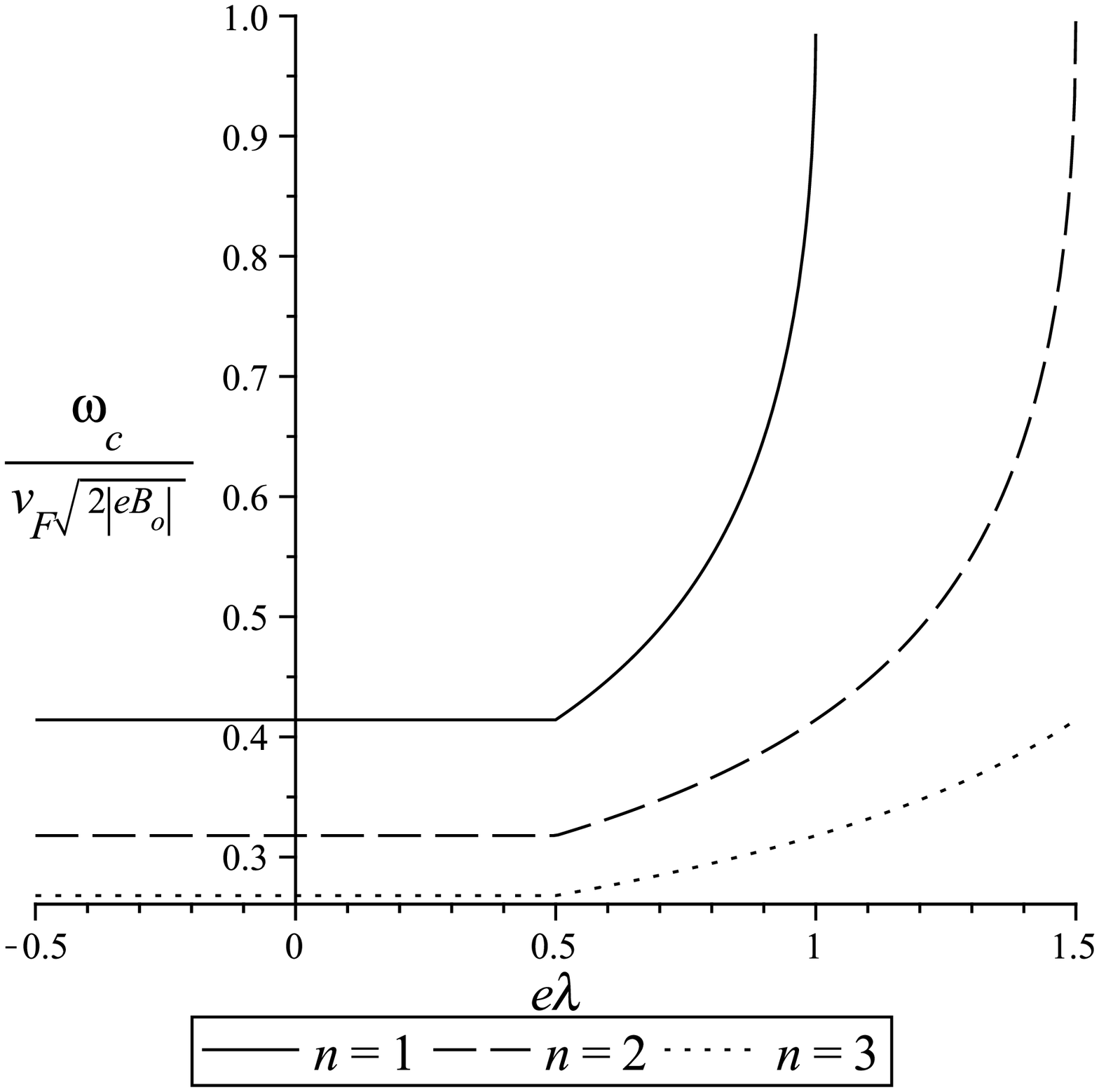}
\vspace{0.5cm} %
b)\includegraphics[height=8cm]{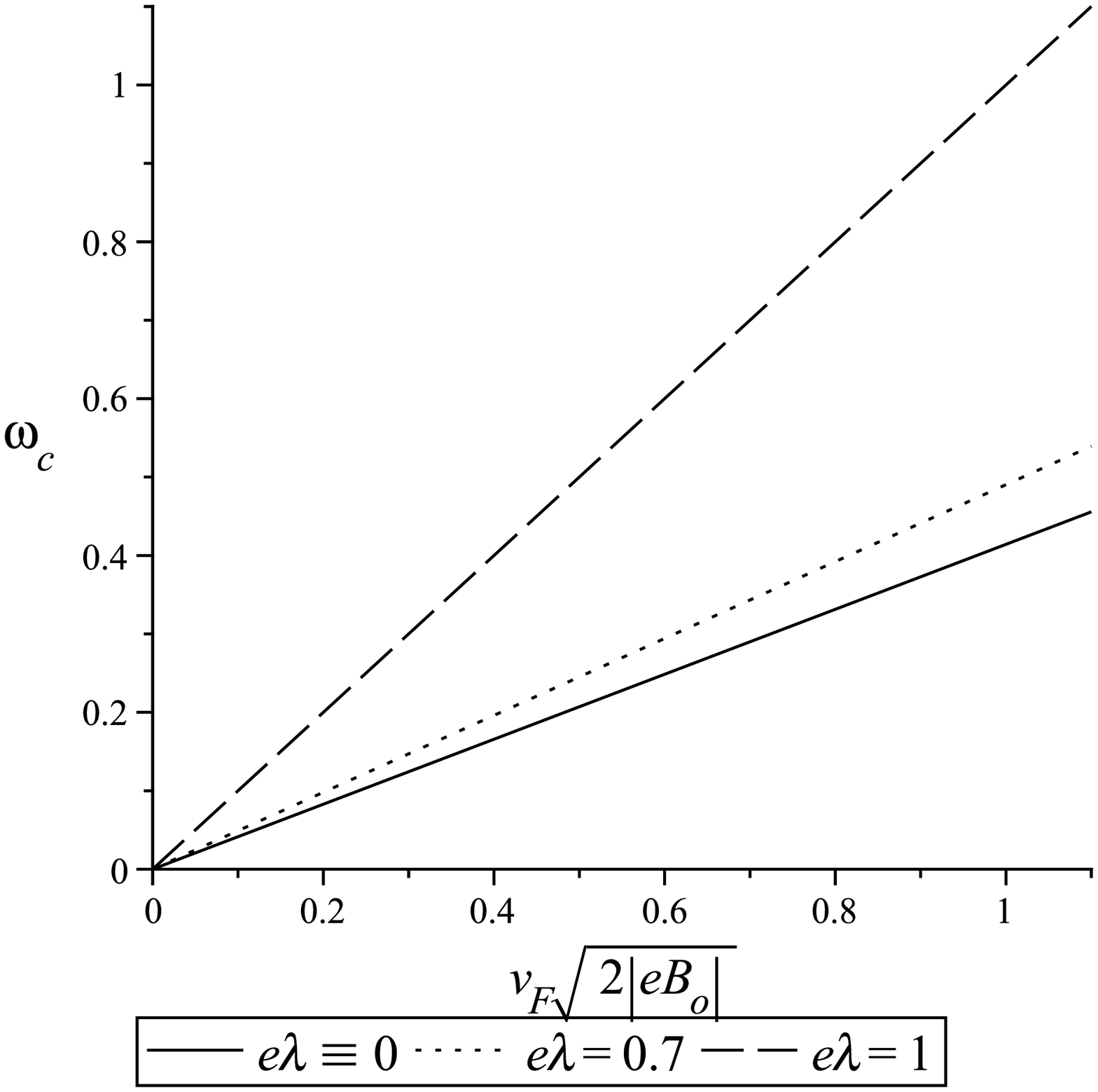}
\caption{\textbf{a)} {\protect\scriptsize Plot of cyclotron frequency $%
\protect\omega _{c}$ versus $e\protect\lambda $. $\protect\omega _{c}$
increases as we raise the parameter $e\protect\lambda \in (1/2,3/2)$. It
does not change when $-1/2<e\protect\lambda \leq 1/2$. This effect is
stronger for lower values of $n$; \textbf{b)} plot of $\protect\omega _{c}$
versus $v_{F}\protect\sqrt{2\left\vert eB_{o}\right\vert }$ for some values
of $e\protect\lambda $ and for $n=1$. }}
\label{fig6}
\end{figure}
The energy spectrum remains unchanged for $-1/2<e\lambda \leq 1/2$. Notice
that the $n=1$ energy mode assumes real values only until $e\lambda =1$.
These results are going to affect the relativistic cyclotron frequency as
\begin{eqnarray}
&&\omega _{c}=v_{F}\sqrt{(}2\left\vert eB_{0}\right\vert )  \notag \\
&&\times \,\Big(\sqrt{n+1+\frac{1}{2}-\frac{\sqrt{1+4e\lambda (e\lambda -1)}%
}{2}-e\lambda }  \notag \\
&&-\sqrt{n+\frac{1}{2}-\frac{\sqrt{1+4e\lambda (e\lambda -1)}}{2}-e\lambda }%
\Big),
\end{eqnarray}%
which is depicted in Figs. \ref{fig6}a and \ref{fig6}b. From them, we see that, for $n=1$, $%
\omega _{c}$ increases as we raise the parameter $e\lambda $ in the interval
$1/2<e\lambda \leq 1$. After $e\lambda =1$, the frequency is imaginary. For $%
n\geq 2$, $\omega _{c}$ increases as we raise the parameter $e\lambda $ in
the interval $1/2<e\lambda <3/2$ and this effect is stronger for lower
values of $n$. Then, many physical properties on graphene which depends on $%
\omega _{c}$ are going to be influenced by the presence of the varying
magnetic field considered here. For example, it might have some impact in
problems involving transitions between Landau levels induced by external
radiation \cite{prlprl}.

For the solutions around the valley $K^{^{\prime }}$, we just change $%
\lambda $ by $-\lambda $. This way, we have $-3/2<e\lambda <1/2$. For $n=0$,
the zero energy is absent when $-3/2<e\lambda <1/2$. The zero energy
mode exists if $-1/2\leq e\lambda <1/2$, as before. We then conclude that
the pseudo-magnetic field given by $1/x^{2}$ fails to observe the
zero Landau level around both valleys, $K$ and $K^{^{\prime }}$.

Other problems were it should be interesting to investigate the consequences
of spatial modulation on the relativistic Landau levels are the relativistic
version of Schrodinger cat states \cite{p4} and the study of quantum phase
transitions \cite{p5}.

\section{Concluding Remarks}

In this work, we investigated how the relativistic Landau levels are
modified if fermions on graphene are held in the presence of a constant
orthogonal magnetic field together with a spatially varying orthogonal
pseudo-magnetic field. We considered the latter falling as $1/x^{2}$%
. We were able to study this problem analytically since our squared Dirac
equation yielded a differential equation called \textit{Biconfluent Heun
equation}, whose solution is well established and has appeared in many
contexts \cite{vary,eug,aura}, helping addressing different physical
problems analytically as we did here. We have observed that such elastic
field, given by $1/x^{2}$, fails to observe the zero Landau level around
both valleys, $K$ and $K^{^{\prime }}$. The consequence is that a Hall
plateau develops at the filling factor $\nu =0$.

We also examined the energy shift due to the presence of the varying pseudo-magnetic field and we investigated how it influences the
relativistic cyclotron frequency. We saw that irregular wavefunctions and
wavefunctions which do not diverge (they are regular solutions) are present.
We observed that the relativistic Landau Levels are unchanged when $%
-1/2<e\lambda \leq 1/2$. So, since we theoretically described a way to
manipulate the relativistic Landau levels, we hope our work being probed in
the context of graphene, a material which has great potential for electronic
applications.

As a final word, we mention that graphene under different position-dependent
magnetic fields was investigated theoretically in reference \cite{many},
including the magnetic field proportional to $1/x^{2}$ alone. It would also
be interesting to investigate them as pseudo-magnetic fields
combined with a constant magnetic field as we did here. If either
simulations or experiments involving graphene fail to observe the zero
Landau level, the presence of varying pseudo-magnetic fields should
be investigated. Another possibility is the presence of \textit{topological
defects} on a graphene sheet, since their existence also split the zero
energy \cite{jana}.

\section*{Acknowledgments}

This work was supported by the CNPq, Brazil, Grants No. 482015/2013-6 (Universal), No. 476267/2013-7  (Universal), No. 306068/2013-3 (PQ) and FAPEMA, Brazil, Grant No. 00845/13 (Universal).

\end{document}